# The Collaborations among Healthcare Systems, Research Institutions, and Industry on Artificial Intelligence Research and Development

Jiancheng Ye[1]*, Michelle Ma[1,2], Malak Abuhashish[1]

[1] Weill Cornell Medicine, Cornell University, New York, NY, USA

[2] Macaulay Honors College in Hunter College, New York, NY, USA

**Corresponding author**

Jiancheng Ye, PhD
Weill Cornell Medicine, New York, NY, USA
jiancheng.ye@u.northwestern.edu

# ABSTRACT

**Objectives:**
The integration of Artificial Intelligence (AI) in healthcare promises to revolutionize patient care, diagnostics, and treatment protocols. Collaborative efforts among healthcare systems, research institutions, and industry are pivotal to leveraging AI's full potential. Understanding these dynamics is essential for addressing current challenges and shaping future AI development in healthcare. This study aims to characterize collaborative networks and stakeholders in AI healthcare initiatives, identify challenges and opportunities within these collaborations, and elucidate priorities for future AI research and development.

**Methods:**
This study utilized data from the Chinese Society of Radiology and the Chinese Medical Imaging AI Innovation Alliance. A national cross-sectional survey was conducted in China (N = 5,142) across 31 provincial administrative regions, involving participants from three key groups: clinicians, institution professionals, and industry representatives. The survey explored diverse aspects including current AI usage in healthcare, collaboration dynamics, challenges encountered, and research and development priorities.

**Results:**
Findings reveal high interest in AI among clinicians, with a significant gap between interest and actual engagement in development activities. Key findings include limited establishment of AI research departments and scarce interdisciplinary collaborations. Despite the willingness to share data, progress is hindered by concerns about data privacy and security, and lack of clear industry standards and legal guidelines. Future development interests focus on lesion screening, disease diagnosis, and enhancing clinical workflows.

**Conclusion:**
This study highlights an enthusiastic yet cautious approach toward AI in healthcare, characterized by significant barriers that impede effective collaboration and implementation. Recommendations emphasize the need for AI-specific education and training, secure data-sharing frameworks, establishment of clear industry standards, and formation of dedicated AI research departments.

**INTRODUCTION**

The advent of Artificial Intelligence (AI) in healthcare represents a paradigm shift, promising unprecedented advancements in medical diagnostics, patient care, and treatment methodologies.[1] AI integration into healthcare is a multifaceted endeavor extending beyond sophisticated algorithm development or cutting-edge technology deployment.[2] It encompasses a broad spectrum of activities including diagnostic imaging enhancements, personalized medicine tailoring, predictive analytics for patient outcomes, and automation of clinical decision-making processes.[3, 4] These applications are transforming the conceptual framework of healthcare delivery while setting new benchmarks for efficiency, accuracy, and patient-centric care.[5] AI technology integration into healthcare is not merely a technological leap; it is an intricate process requiring harmonious collaborative efforts among various stakeholders.[6]

Essential to harnessing AI's full potential in healthcare is strategic collaboration and synergy among key stakeholders: healthcare practitioners who contribute clinical expertise and insights, research institutions that drive innovation through rigorous scientific inquiry, and industry stakeholders that translate technological advancements into viable healthcare solutions.[7] This tripartite collaboration is pivotal for overcoming translational hurdles that often hinder seamless AI technology integration into clinical settings, ensuring that AI innovations are not only technologically robust but also aligned with practical healthcare needs, making them readily adoptable in real-world scenarios.[8]

However, fostering effective collaboration among these diverse entities presents significant challenges. Disparate objectives, varying operational cultures, and rapid AI technology evolution often lead to misalignments that can impede collaborative efforts.[9] Additionally, regulatory considerations, ethical concerns, and data privacy issues add complexity layers to these partnerships.[10] Navigating these challenges requires deep understanding of the collaborative ecosystem, clear articulation of common goals, and establishment of frameworks that facilitate mutual engagement and benefit sharing.

To unravel collaboration intricacies in the AI healthcare domain, this study relies on data from the Chinese Society of Radiology and the Chinese Medical Imaging AI Innovation Alliance.[11] These organizations serve as repositories of valuable insights, encapsulating experiences and perspectives of clinicians, researchers, and industry professionals. Through a meticulously designed national cross-sectional survey, we aim to illuminate collaborative relationships that define the current state of AI in healthcare. This paper seeks to achieve several key objectives: (1) explore existing networks and partnerships among clinicians, research institutions, and industry stakeholders engaged in AI healthcare initiatives; (2) uncover challenges faced by stakeholders in AI adoption and implementation in healthcare while identifying opportunities for overcoming these hurdles; (3) provide insights into priorities and areas of interest for future research and development in the AI healthcare domain.

This study examines the intricate dynamics of collaboration among healthcare systems, research institutions, and industry stakeholders, providing a comprehensive overview of the current landscape, inherent challenges, and prospective directions in AI research and implementation within healthcare.

**METHODS**

Data utilized in this study were obtained from publicly available datasets released by two prominent Chinese organizations: the Chinese Society of Radiology and the Chinese Medical Imaging AI Innovation Alliance.[11] The data are publicly available and do not require Institutional Review Board approval. These datasets compile comprehensive information related to AI research and implementation in healthcare, with specific emphasis on radiology and medical imaging. Survey participants were meticulously selected to ensure diverse perspectives from key stakeholders in the AI healthcare ecosystem. The study included individuals from three distinct groups: clinicians, researchers from renowned healthcare research institutions, and representatives from the healthcare industry. This stratification was essential to capture a holistic view of the collaborative landscape among healthcare systems, research institutions, and industry players. Descriptive statistics were performed to derive meaningful insights from survey responses.

The clinician group comprised healthcare professionals actively engaged in clinical practice, including radiologists, physicians, and other specialists. Their firsthand patient care experience provided valuable insights into practical applications and challenges of AI technologies in clinical settings. Research institutions group participants were affiliated with prominent healthcare research institutions across China. These individuals brought extensive knowledge and expertise in innovative research and development within the AI healthcare domain, contributing to academic and scientific aspects of the study. The industry group consisted of professionals from companies and organizations actively involved in development, deployment, and commercialization of AI solutions in healthcare. This group's perspective illuminated technological advancements, market trends, and industry needs in the AI healthcare sector.

To gather comprehensive and relevant information, a structured survey instrument was designed collaboratively by experts in healthcare, AI, and survey methodology. The survey questionnaire encompassed a range of topics including: (1) current AI usage in healthcare—examining current integration of AI technologies in healthcare practices; (2) challenges and barriers—identifying challenges faced by clinicians, researchers, and industry professionals in AI adoption and implementation in healthcare; (3) collaboration dynamics—investigating existing collaborations, partnerships, and communication channels among healthcare systems, research institutions, and industry stakeholders in the AI domain; and (4) research and development priorities—understanding key areas of interest and priorities for future research and development in AI healthcare.

**Study Approval**

We used exclusively publicly available data for this study.

**RESULTS**

**Clinicians group**

<u>Characteristics of participants</u>

In our study, we distributed 5,148 questionnaires within the clinician group, achieving a 99.9% response rate (5,142 responses). These surveys spanned 31 regions nationwide, incorporating 2,135 hospitals. The most represented age group among clinicians was 30-40 years (34%), closely followed by 40-50 years (33%). This age distribution was consistent across both secondary and tertiary hospitals.

Respondents' educational qualifications provide valuable insights into their academic backgrounds. The predominant qualification was a bachelor's degree (58%), followed by master's degrees (22%) and professional or doctorate degrees (20%). Notably, half of the clinicians, especially those with keen interest in AI, were attending physicians or deputy chief physicians with over 15 years of radiology experience. Among these, 66% worked in secondary hospitals and 50% in tertiary hospitals.

Professional roles varied, with 27% of participants serving as department directors and 13% as deputy department heads. Primary clinical research areas were diverse, focusing on abdomen (56%), chest (45%), bone and joints (36%), nervous system (35%), and head and neck (29%), with additional interests in breast, pediatric, interventional radiology, and molecular imaging.

Current state of health Information systems and AI Collaboration

**Table 1** presents the current state of health information systems and AI collaboration in the clinician group. Overall, 47% of hospitals lack structured imaging report systems, 43% are planning implementation, and only 10% have established and actively use such systems. These systems are predominantly applied in departments addressing lung nodules or lung cancer, colorectal cancer, breast cancer, and coronary artery disease. Regarding hospital informatization, 63% of patient information is accessible through unified systems, 31% requires querying across multiple systems, and 6% remains inaccessible.

In terms of AI involvement, 74% of clinicians have not engaged in AI-related research. Among the remainder, 21% participated without producing results, 4% published research papers, 0.8% developed AI products, 0.5% secured related patents, and only 0.4% received domestic or international recognition. Additionally, 84% of clinicians have not collaborated with relevant enterprises or research institutions, and 92% have not worked with imaging device companies.

Infrastructure for AI research

The majority of hospitals (72%) lack departments dedicated to AI research. A significant portion of clinicians (27%) are uncertain about such departments' existence within their institutions, with only 1% of hospitals having established specialized AI research departments. Within these facilities, 55% do not have departments focused on translating research findings into practical applications. Only 20% have established such departments, while 25% of clinicians are unsure of their presence. Tertiary hospitals exhibit comparatively more robust AI research infrastructure, with 27% hosting relevant departments, contrasting sharply with 8% in secondary hospitals. A concerning 79% of hospitals lack engineering personnel engaged in AI research. Of the limited departments that exist, 11% have 1-2 individuals with at least master's degrees, and only 4% have staff of five or more with comparable qualifications.

Data sharing for AI research

**Table 2** shows that clinicians primarily contribute to AI research collaborations through image data sharing (89%), clinical information provision (76%), assistance with image data labeling (70%), identification of clinical needs and issues (70%), and feedback on AI products (52%). The preferred AI collaboration method involves complimentary data sharing for joint research publications or patents, favored by 55% of respondents, followed by free data sharing in exchange for AI product discounts (25%), and intra-hospital sharing or paid purchases (20%). Regarding data privacy and security, a majority (74%) recognize the need for data anonymization and confidentiality agreements, while 23% are unsure of relevant policies, and 3% find these measures unnecessary.

Despite widespread awareness, 74% of clinicians have only heard of AI-related products without actual usage. In contrast, 20% have utilized such products, 5% are engaged in AI product development, and only 1% have contributed to product development with concrete outcomes. AI products find their most significant application in lung nodule screening, utilized by 88% of hospitals, with coronary artery analysis (6%) and other areas like bone age, breast, and prostate diagnostics trailing in usage.

Challenges in the AI collaboration

**Table 3** highlights challenges and barriers encountered in current AI collaborations. A significant 65% of clinicians identify absence of industry standards as a principal obstacle in AI research, while 63% cite lack of legal guidelines for employing AI products in clinical tasks. Moreover, 59% of clinicians report gaps in relevant AI knowledge. Concerns about AI product credibility and extensive workforce requirements are noted by 56% and 45% of respondents, respectively. Notably, 56% of clinicians express concerns that AI could lead to misdiagnoses or missed diagnoses, potentially resulting in critical medical errors. While 27% fear AI underperformance or failure, only 13% foresee no negative impacts on healthcare systems.

Future development in clinicians group

**Table 4** presents prospects for future development in healthcare AI as perceived by the clinician group. A vast majority (90%) of clinicians anticipate needing substantial exploration time. Meanwhile, 25% are skeptical about achieving short-term practical results, 8% predict eventual replacement of radiologists, and 3% dismiss AI as a passing fad without practical utility. The primary interest area for healthcare AI among clinicians is lesion screening and detection (84%), followed by disease diagnosis (65%), and prognosis analysis and treatment effectiveness evaluation (64%). Medical education emerges as another significant interest area (41%). Clinicians also show enthusiasm for collaboration beyond research institutions, especially with industry (50% support) and forming internal AI teams (25%).

Clinicians express strong preference for research institutions or technology companies to supply AI devices and software for clinical trials (88%), develop image processing algorithms (73%), and provide research and funding support (60%). A small fraction (0.3%) seeks additional resources like training and education opportunities. Most clinicians (93%) anticipate research output timelines exceeding one year, with expectations divided between one to two years (43%) and over two years (50%). Only 7% expect results within one year.

Recommendations from clinician group

Clinicians prioritize support through various avenues: developing collaborative platforms with AI companies (82%), bringing expert research teams (64%), and conducting regular training workshops to disseminate AI knowledge (52%). They propose key recommendations to advance healthcare AI development: (1) elevating training and knowledge levels, ensuring expertise reaches grassroots hospitals for direct clinical application; (2) establishing platforms for multi-center cooperation, enabling efficient resource sharing and communication among healthcare providers; (3) enhancing accuracy and usability of AI products, striving to significantly reduce misdiagnosis or missed diagnosis risks; (4) creating comprehensive industry standards, legal frameworks, and dispute resolution mechanisms, ensuring uniform AI approaches in healthcare; and (5) implementing standardized data protection management practices, safeguarding patient information against unauthorized access.

**Research institutions group**

A total of 120 surveys from research institutions across 19 regions nationwide were collected and analyzed. Age distribution shows a youthful skew: 69% under 30 years, 18% between 30-40 years, 10% between 40-50 years, and 3% over 50 years. Educationally, the group is highly qualified, with postgraduates comprising 58% and doctoral candidates, undergraduates, and college graduates making up the remainder. Notably, postgraduates and doctoral candidates together represent 91% of respondents. Healthcare AI research team sizes vary: 30% have 1-2 members, 28% have teams larger than 10, 23% work in groups of 6-10, and 19% operate in teams of 3-

5. Leadership positions include 13% as research group heads, 6% as academic or institutional lab directors, and 2% leading provincial or ministerial key labs.

Current research status

**Table 5** illustrates research status in the research institutions group. Predominant research areas are image classification/segmentation/target detection (42%), video image analysis (40%), and molecular imaging (36%). Other areas such as imaging methods, image reconstruction algorithms, reinforcement learning, and biometric identification each attract 12% focus. Less common fields include control systems and engineering, natural language processing, and autonomous driving, each with 5% share. Application-wise, lung nodule screening (32%), pathological diagnosis (25%), and early tumor diagnosis (25%) are most frequent, followed by breast disease screening (20%), stroke diagnosis (17%), and coronary heart disease diagnosis (12%). Emerging areas include retinal lesion screening, fracture screening, and bone age detection.

Primary research focus is assisting diagnosis and clinical treatment decisions (66%), with significant emphasis on enhancing clinical workflows and optimizing AI imaging methods (48%). Additionally, data security remains a concern for 18% of researchers. Regarding achievements, the majority resulted in scientific publications, with journal papers (47%) and conference papers (37%) leading. Additionally, 26% of researchers secured AI product patents, and 7% received domestic or international awards. Despite this, over 70% of researchers have not contributed to healthcare AI product development, with only 30% engaging in such projects.

Collaborations with healthcare systems

**Table 6** provides an overview of collaborative landscape between research institutions and healthcare systems. One-third (33%) of researchers collaborated with just one hospital, while 31% have partnerships with 2-5 hospitals. Notably, another 31% have not engaged in prior collaborations. Collaborations extending to 6-10 hospitals or beyond are relatively rare, cumulatively accounting for only 5%. An important hurdle lies in obtaining meaningful data, as over one-third (33%) of researchers report not acquiring valuable data from collaborations. When data is accessible, it predominantly encompasses sample sizes from tens to hundreds of patients (22% and 26%, respectively). Only a minority access larger datasets, with 13% obtaining data from thousands of patients and 6% accessing data from over ten thousand patients.

Half of researchers (50%) currently do not collaborate with any healthcare systems. Among those who do, the vast majority (39%) work with fewer than five systems. A smaller fraction engages with 5-10 systems (12%), and only 2% have partnerships with over ten healthcare systems. Collaboration with relevant companies is also limited; nearly half report no such collaborations. Among those who collaborate, 43% work with fewer than five companies. Collaborations with 5-10 companies and more than ten companies represent smaller portions, totaling 10%.

Data Sharing and Data Security

Table 7 demonstrates data sharing and security in the research institutions group. Researchers exhibit clear preferences for data sharing mechanisms. A majority (55%) prefer sharing data without cost, resulting in co-authored research papers or patents, followed by 25% who prefer free data sharing coupled with AI product discounts, and 20% who engage in data sharing within hospital networks or paid purchases. Regarding data privacy and security, 74% of researchers affirm the necessity of data anonymization and confidentiality agreements. However, 23% remain uncertain about relevant policies, and 3% view these precautions as unnecessary.

Among researchers, free data sharing for co-authored outputs emerges as most popular, capturing 72% approval, followed by 19% who prefer free data sharing with AI product discounts, and 9% who favor paid data acquisition. Despite recognizing data security importance, 30% of researchers admit uncertainty about achieving it effectively. Another 30% underscore the critical need for understanding and implementing robust data security measures, emphasizing mastering key technologies essential for data protection. Nevertheless, 20% of respondents express no specific data security concerns, and only 15% possess both critical technologies for data protection and practical implementation experience.

Infrastructure for AI research

Table 8 outlines AI research infrastructure state across research institutions. Nearly half (47%) are establishing AI infrastructure, while a similar proportion (46%) plan to but have not yet initiated. A small minority (7%) have successfully established and currently utilize AI systems, mainly focusing on structured image reports for lung cancer and other tumor-related diseases.

Regarding dedicated AI research departments or organizations establishment, there is notable lack of awareness and implementation among researchers. Forty-five percent are uncertain about such entities' existence within their institutions. Similarly, 45% report no dedicated AI research department or organization exists, with only 10% confirming their presence. Concerning capability for transforming research outcomes into practical applications, opinions are divided. Forty-five percent acknowledge relevant departments' presence within their institutions. However, the majority (55%) indicate lack of such departments, with 41% unsure of their existence and 14% explicitly stating their absence.

Current state of AI research and development

**Table 9** provides insights into AI research and development landscape as perceived by respondents. A significant majority (64%) have familiarity with AI-related products through hearsay rather than direct usage. Meanwhile, 18% are actively engaged in ongoing AI research and development efforts. A smaller fraction (14%) has hands-on experience with AI products, and only 4% have contributed to research and development yielding tangible outcomes for the AI research community.

Research-to-product application transition presents notable hurdles. Over half (53%) highlight the substantial gap between research achievements and their conversion into practical applications as a primary challenge, underscoring difficulties in achieving rapid market readiness. Similarly, 52% point out intensive workload and significant human resource requirements needed for AI research and development. Other challenges include lack of industry standards (4%), questions surrounding AI product credibility (18%), unclear legal responsibilities between products and clinical practitioners (3%), and absence of relevant knowledge and perseverance (18%). Furthermore, 37% identify the main challenge as fostering effective collaboration among healthcare systems, research institutions, and industry. Data acquisition and processing are viewed as the most daunting challenge by 29%, overshadowing concerns related to capital investment (14%), algorithm support (12%), and policy facilitation (7%).

Future development in the research institutions group

**Table 10** demonstrates development prospects of AI in healthcare from the research institutions group. A vast majority (86%) see significant value in further exploring AI products for healthcare, although 28% temper expectations with caution about short-term viability of practical applications. A minority (7%) view AI as a fad lacking practical utility, while 5% anticipate that AI will eventually supersede traditional imaging methods. Interest among researchers is notably high in lesion screening and detection (71%), disease diagnosis and prognosis analysis (66%), evaluating treatment effectiveness (58%), and medical education (36%).

Collaborative endeavors are highly sought after, with 75% keen on partnering with healthcare systems and 71% advocating for increased collaboration between research institutions. Industry and educational institution collaborations attract support rates of 51% and 27%, respectively, with 17% preferring to develop their own teams. Regarding resources, researchers prioritize access to scientific publications (65%) and research grants (40%). Other valuable resources include algorithm research (30%), product development (19%), and product usage training (19%). For hospital collaborations, researchers express strong desire for access to imaging and clinical data (85%), image annotation (56%), support in applying for research projects (53%), software transformation assistance (47%), clinical project determination knowledge (46%), and research funding (39%).

Expected outcomes from AI collaborations are predominantly research papers (67%), followed by patents (62%), AI product developments (56%), research funding (50%), and personal achievements (26%). Regarding timeline, 80% anticipate research outputs will materialize over one year, with 43% expecting results within 1-2 years and 37% foreseeing outcomes beyond two years. A smaller group (20%) hopes for results within the first year. Most coveted support from collaborators includes establishing platforms for cooperation with AI companies (69%), integrating hospital projects and data into research (66%), and organizing regular workshops for knowledge exchange (64%).

**DISCUSSION**

This study provides comprehensive examination of current state and challenges of collaboration among healthcare systems, research institutions, and industry in AI research and development. Findings depict a landscape marked by substantial interest but hindered by notable barriers to effective collaboration and implementation.

The impressive clinician response rate highlights strong interest in AI's potential to transform healthcare. Yet, the stark contrast between this interest and actual engagement in AI-related research or development activities indicates systemic barriers. The fact that most hospitals lack structured imaging report systems—basic yet crucial infrastructure for AI integration—underscores a fundamental readiness gap for AI adoption across healthcare institutions. This gap is further evidenced by limited establishment of dedicated AI research departments, which are crucial for fostering innovation and translating research into clinical practice.[12, 13]

The demographic and professional profile of clinician respondents, predominantly those with bachelor's degrees and substantial radiology experience, suggests a workforce theoretically well-positioned to contribute to AI advancements. However, minimal involvement in AI research or development activities, alongside scant production of tangible outcomes such as patents or internationally recognized research, indicates a disconnect between potential and actualized contribution. This disconnect may stem from reported lack of AI-specific knowledge and training, highlighting a critical intervention area.[14]

Willingness among clinicians to share data for AI research, coupled with recognition of data anonymization and confidentiality importance, presents an opportunity for leveraging clinical insights for AI development. However, actual collaboration mechanisms and existing concerns around data privacy and security protocols suggest that more structured, transparent, and secure frameworks for data sharing are necessary to fully realize this potential.[15]

Findings regarding AI collaboration challenges, particularly absence of industry standards and legal guidelines, resonate with broader issues facing AI in healthcare. These challenges, along with reported gaps in relevant AI knowledge among clinicians, underscore the need for multifaceted approaches to address these barriers. Future healthcare AI developments, as anticipated by clinicians, emphasize lesion screening and detection, disease diagnosis, and prognosis analysis, pointing toward areas where collaboration between healthcare systems, research institutions, and industry could be most fruitful.[16]

The research institutions perspective complements and expands understanding of the collaboration landscape.[17] Focus on image classification, segmentation, and analysis among research institutions aligns with clinicians' interests in diagnostic AI applications.[18] However, challenges in acquiring impactful data and limited collaboration with healthcare systems highlight systemic barriers to effective data exchange and utilization for AI development.

To navigate the intricate landscape outlined by our findings and effectively harness AI potential within healthcare, several strategic recommendations emerge as pivotal. First, addressing the critical gap in AI-specific knowledge among clinicians is paramount. This involves developing and disseminating targeted education and training programs that not only familiarize healthcare professionals with AI technologies but also equip them with skills to actively participate in AI research and development.[19] Such initiatives could be spearheaded through collaborative efforts between educational institutions, healthcare systems, and industry partners, ensuring curriculum is both comprehensive and applicable to current clinical practices.[20]

Second, establishing robust, secure data-sharing frameworks is essential. These frameworks should prioritize patient privacy and data security while facilitating seamless information exchange between healthcare systems and research entities. By implementing standardized protocols and leveraging advanced encryption technologies, these frameworks can alleviate data privacy concerns and enhance collaboration efficiency.[21] Furthermore, these systems should be designed for interoperability, allowing integration of diverse data sources and thereby enriching data available for AI research.[22]

Third, absence of clear industry standards and legal guidelines has been identified as a significant barrier to AI integration in healthcare. Thus, there is pressing need for regulatory bodies, in collaboration with healthcare professionals, researchers, and industry stakeholders, to develop comprehensive standards and guidelines.[23] These regulations should address ethical considerations, data usage, and AI technology deployment, ensuring AI applications in healthcare are both safe and effective.[24, 25] Moreover, establishing legal frameworks can help clarify all parties' responsibilities and protect patient interests, thereby fostering a more trustworthy environment for AI development.

Establishing dedicated AI research departments within healthcare institutions represents another vital step toward bridging the gap between potential and actualized AI advancements. These departments could serve as innovation hubs, facilitating AI research translation into clinical applications.[26, 27] By fostering closer collaboration between clinicians and AI researchers, these departments can ensure AI developments align with clinical needs and are rapidly integrated into healthcare practices.[28] Fostering multidisciplinary collaborations stands out as a vital strategy for advancing AI in healthcare.[29] By bringing together expertise of healthcare professionals, AI researchers, and industry innovators, these collaborations can drive development of AI applications that are not only technologically advanced but also deeply attuned to clinical care complexities.[30, 31] Such partnerships should aim to leverage unique strengths and perspectives of each sector, ensuring AI technologies are developed in ways that are both innovative and grounded in real-world healthcare needs.[32]

**Limitation**

This study has several limitations. First, reliance on survey data, although providing valuable insights from a wide range of respondents, carries inherent limitations related to self-reporting, including potential response biases and accuracy of self-assessed knowledge and experiences. Despite efforts to ensure comprehensive and diverse respondent pools, findings may not fully capture the breadth of perspectives across all healthcare settings and geographical regions. Second, the study's focus on China, while offering in-depth insights into collaborative landscape within a major healthcare and technological market, might limit findings' generalizability to other contexts. Different countries may have unique regulatory environments, technological infrastructures, and cultural attitudes toward AI in healthcare, which could influence collaborative efforts' nature and success in ways not captured by this study. Finally, the study's quantitative approach, while effective for identifying broad trends and patterns, may not fully capture collaborative relationship complexities, interdisciplinary communication nuances, or qualitative aspects of innovation and problem-solving in AI research and development. Future research could benefit from incorporating qualitative methods such as interviews or case studies to gain deeper insights into these aspects.

**CONCLUSION**

This study illuminated the complex and dynamic landscape of collaboration among healthcare systems, research institutions, and industry stakeholders in the AI domain. The enthusiastic response from clinicians underscores widespread recognition of AI's potential to enhance patient care, improve diagnostic accuracy, and streamline healthcare operations. However, this enthusiasm is tempered by implementation challenges, including lack of structured imaging report systems, insufficient collaboration between key stakeholders, and dearth of dedicated AI research and development departments within hospitals. These challenges are further compounded by concerns over data privacy and security and absence of clear industry standards and legal guidelines, which collectively hinder seamless AI technology integration into clinical practice. The recommendations emanating from this study underscore the critical need for concerted efforts by all stakeholders to address identified challenges. By fostering collaborative ecosystems that encourage sharing of knowledge, resources, and expertise, we can accelerate the pace of AI innovation and its healthcare applications. Moreover, establishing comprehensive industry standards and legal frameworks will provide necessary foundations for building trust and credibility in AI technologies, ensuring their ethical and effective use in patient care.

**FUNDING**

None.

**CONTRIBUTION STATEMENT**

JY designed the study, contributed to the data analyses, and wrote the manuscript. MM contributed to data analyses. All authors read and approved the final version of the manuscript.

**DATA AVAILABILITY STATEMENT**

All data are incorporated into the article and its online supplementary material.

**CONFLICT OF INTEREST STATEMENT**

None.

**Tables/Figures**

**Table 1.** Current state of AI collaboration in healthcare systems from the clinician group

| Characteristics | Percentage |
| --- | --- |
| **Structured imaging report system** | |
|     Not established | 47% |
|     Planning to establish | 43% |
|     Already established and operational | 10% |
| **Hospital informatization level** | |
|     Accessible within unified system | 63% |
|     Requires queries in different system | 31% |
|     Not accessible | 6% |
| **Involvement in AI-related research** | |
|     Not involved | 74% |
|     Participated but no results | 21% |
|     Published research papers | 4% |
|     Developed AI products | 0.8% |
|     Obtained related patents | 0.5% |
|     Research received domestic or international awards | 0.4% |
| **Clinician collaboration with research institutions** | |
|     No collaboration with enterprises or research institutions | 84% |
|     No collaboration with imaging device companies | 92% |

**Table 2.** Data sharing and collaboration mechanisms in the clinician group

| Characteristics | Percentage |
|---|---|
| **Resources for AI collaboration from clinicians** | |
|     Image data sharing | 89% |
|     Clinical Information | 76% |
|     Image data labeling assistance | 70% |
|     Clinical needs and issues | 70% |
|     AI Product feedback | 52% |
| **AI collaboration mechanisms** | |
|     Free data sharing for research papers or patent | 55% |
|     Free data sharing with AI product discounts | 25% |
|     Sharing within hospital or paid purchasing | 20% |
| **Opinions on desensitizing data and signing confidentiality agreements** | |
|     Necessary | 74% |
|     Unclear about relevant policies | 23% |
|     Unnecessary | 3% |
| **Clinicians' engagement with AI-Related Products** | |
|     Aware of AI-related products but haven not utilized them | 74% |
|     Have used AI-related products | 20% |
|     Actively involved in product development | 5% |
|     Participated in product development with tangible results | 1% |

**Table 3.** Challenges in the AI collaboration in the clinician group

| Characteristics | Percentage |
| --- | --- |
| **Perceived problems in AI research** | |
|     Lack of industry standards | 65% |
|     Lack of legal guidelines | 63% |
|     Lack of relevant knowledge in the AI field | 59% |
|     Low credibility of AI products | 56% |
|     Significant workforce requirements | 45% |
| **Beliefs about AI's impact on healthcare systems** | |
|     May lead to misdiagnosis or missed diagnosis | 56% |
|     May perform poorly or fail to work | 27% |
|     Will have no negative impact on healthcare systems | 13% |

**Table 4.** Future Development in the clinician group

| Characteristics | Percentage |
| --- | --- |
| **Opinions on prospects of AI research** | |
|     Requires considerable time for exploration | 90% |
|     Practical results will not be achieved in the short term | 25% |
|     Radiologists will eventually be replaced | 8% |
|     Considered a passing trend with no practical value | 3% |
| **Interests in healthcare AI fields** | |
|     Lesion screening and detection | 84% |
|     Disease diagnosis | 65% |
|     Disease prognosis analysis and treatment effectiveness evaluation | 64% |
|     Medical education | 41% |
| **Expectations for collaboration** | |
|     Collaboration with research institutions | 88% |
|     Collaboration with enterprises | 50% |
|     Forming AI teams | 25% |
| **Desired resources for collaboration** | |
|     AI equipment and software | 88% |
|     Construction of image processing algorithms | 73% |
|     Research and funding support | 60% |
|     Training and education opportunities | 0.3% |
| **Expected time frame for research output** | |
|     Within one year | 7% |
|     Within one to two years | 43% |
|     Over two years | 50% |

**Table 5.** Research status in the research institutions group

| Characteristics | Percentage |
|---|---|
| **Research directions** | |
| Image classification/segmentation/target detection | 42% |
| Video image analysis | 40% |
| Molecular imaging | 36% |
| Imaging methods | 12% |
| Imaging reconstruction algorithm | 12% |
| Reinforcement learning | 12% |
| Biometric identification | 12% |
| Control systems and control engineering | 5% |
| Natural language processing | 5% |
| Autonomous driving | 5% |
| **Common application areas** | |
| Lung nodule screening | 32% |
| Pathological diagnosis | 25% |
| Early tumors Diagnosis | 25% |
| Breast disease screening | 20% |
| Stroke diagnosis | 17% |
| Coronary heart disease auxiliary diagnosis | 12% |
| **Primary issues addressed using AI technology** | |
| Auxiliary diagnosis and clinical treatment decision-making | 66% |
| Streamlining workflows and optimizing AI imaging methods | 48% |
| Improving data security | 18% |
| **Achievements in AI-related research** | |
| Journal papers | 47% |
| Conference papers | 37% |
| Obtained AI product patents | 26% |
| Received domestic or international awards | 70% |
| Involved in the AI research | 30% |

**Table 6.** Collaborations between research institutions and healthcare systems from the research institutions group

| Characteristics | Percentage |
|---|---|
| **Number of collaborative hospitals** | |
| No prior collaboration | 31% |
| Collaborate with 1 hospital | 33% |
| Collaborate with 2-5 hospitals | 31% |
| Collaborate with > 5 hospitals | 5% |
| **Access to impactful data** | |
| No impactful data | 33% |
| Tens to hundreds of cases | 22% |
| Hundreds to thousands of cases | 26% |
| Several thousand cases | 13% |
| Over ten thousand cases | 6% |
| **Collaboration with industry** | |
| Do not collaborate with any healthcare systems | 50% |
| Collaborate with <5 healthcare systems | 39% |
| Collaborate with 5-10 healthcare systems | 12% |
| Collaborate with >10 healthcare systems | 2% |
| **Collaboration with relevant companies** | |
| Do not have collaborations with any relevant companies | 47% |
| Collaborate with ≤ 5 companies | 43% |
| Collaborate with >5 companies | 10% |

**Table 7.** Data sharing and data security in the research institutions group

| Characteristics | Percentage |
|---|---|
| **Data sharing methods in AI collaborations** | |
|     Free data exchange with sharing of research papers | 72% |
|     Free data sharing with discounted access to AI products | 19% |
|     Paid acquisition | 9% |
| **Concerns about data security** | |
|     Express importance of data security but lack certainty on how to ensure it | 35% |
|     Consider it crucial to understand and maintain processes for data security | 30% |
|     Show no particular concern about data security issues | 20% |
|     Possess key technologies for maintaining data security along with practical experience | 15% |

**Table 8.** Infrastructure for AI research in the research institutions group

| Characteristics | Percentage |
| --- | --- |
| **Laboratory status of establishing structured imaging reports** | |
|     Preparing to establish | 47% |
|     Intend to initiate but have not yet started | 46% |
|     Already established and in active use | 7% |
| **Researchers' awareness of AI research institutions or organization** | |
|     Unaware of establishment status | 45% |
|     No AI research institution or organization established | 45% |
|     Aware of existing AI research institutions and organizations | 10% |
| **Researchers' awareness of departments for transforming results into practical applications** | |
|     Institution has relevant departments | 45% |
|     Lacks information about such departments | 41% |
|     No such department exists | 14% |

**Table 9.** Current state of AI research and development in the research institutions group

| Characteristics | Percentage |
| --- | --- |
| **Experience with AI-Related Products** | |
| Aware of AI-related products but have not utilized them | 64% |
| Active involvement in ongoing research and development | 18% |
| Have used related products | 14% |
| Participated in research and development with relevant results | 4% |
| **Challenges in AI research process** | |
| Difficulty transitioning into practical applications | 53% |
| Substantial workload and manpower investment | 52% |
| Lack of industry standards | 4% |
| Credibility of AI products | 18% |
| Unclear legal responsibilities between products and clinical practitioners | 3% |
| Absence of relevant knowledge and perseverance | 18% |
| Collaboration among healthcare systems, research institutions, and industry | 37% |
| Data acquisition and processing | 29% |
| Capital injection | 14% |
| Algorithm support | 12% |
| Policy support | 7% |

**Table 10.** Future development in the research institutions group

| Characteristics | Percentage |
| --- | --- |
| **Prospects of AI research** | |
| Worthy of investigation | 86% |
| Practical results not expected short term | 28% |
| Viewed as a temporary trend | 7% |
| Traditional methods may be replaced | 5% |
| **Areas of interest in AI** | |
| Lesion screening and detection | 71% |
| Disease diagnosis and prognosis analysis | 66% |
| Treatment effectiveness | 58% |
| Medical education | 36% |
| **Interest in collaboration** | |
| With medical institutions | 75% |
| With research institutions | 71% |
| With enterprises | 51% |
| With educational institutions | 27% |
| Favor building own teams | 17% |
| **Resources expected from collaborators** | |
| Published papers | 65% |
| Research project design | 40% |
| Algorithm research | 30% |
| Product development | 19% |
| Product usage training | 19% |
| **Resources expected from hospitals in collaborations** | |
| Access to imaging and clinical data | 85% |
| Image annotation | 5% |
| Joint application for research projects | 16.5% |
| Assistance with software transformation | 13.4% |
| Knowledge of clinical project determination | 17.4% |
| Research funding | 16.4% |
| **Expected main output of AI collaboration** | |
| Research papers | 67% |
| Patents | 62% |
| AI Products | 56% |
| Research funding | 50% |
| Personal gains | 26% |
| **Expected time frame for research output** | |
| Within one year | 20% |
| Within one to two years | 43% |
| Over two years | 37% |
| **Desired support from collaborators** | |
| Collaboration platform with AI companies | 69% |
| Introducing hospital projects and data for research | 66% |
| Regular training-related workshops | 64% |